\documentclass[aip,jmp,graphicx,reprint]{revtex4-1}
\draft
\usepackage{dcolumn}
\usepackage{graphicx}
\usepackage{subfigure}
\usepackage{epstopdf}
\usepackage{color}

\newcommand*{\citen}[1]{%
  \begingroup
    \romannumeral-`\x % remove space at the beginning of \setcitestyle
    \setcitestyle{numbers}%
    \cite{#1}%
  \endgroup   
}
\begin{document}

\title{Compact tunable Compton x-ray source from laser-plasma accelerator and plasma mirror}

\author{Hai-En Tsai}
% \email{Haien@utexas.edu}
 \affiliation{Department of Physics, The University of Texas at Austin, Austin, Texas 78712-1081, USA}%Lines break automatically or can be forced with \\
 \author{Xiaoming Wang}
 \affiliation{Department of Physics, The University of Texas at Austin, Austin, Texas 78712-1081, USA}

 \author{Joseph M. Shaw}
  \affiliation{Department of Physics, The University of Texas at Austin, Austin, Texas 78712-1081, USA} 
  
  \author{Zhengyan Li}
  \affiliation{Department of Physics, The University of Texas at Austin, Austin, Texas 78712-1081, USA}

  \author{Alexey V. Arefiev}
  \affiliation{Institute for Fusion Studies, The University of Texas at Austin, Austin, Texas 78712, USA}
 
  \author{Xi Zhang}
  \affiliation{Department of Physics, The University of Texas at Austin, Austin, Texas 78712-1081, USA}\affiliation{Institute for Fusion Studies, The University of Texas at Austin, Austin, Texas 78712, USA}
 \author{Rafal Zgadzaj}
 \affiliation{Department of Physics, The University of Texas at Austin, Austin, Texas 78712-1081, USA} \author{ Watson Henderson}
 \affiliation{Department of Physics, The University of Texas at Austin, Austin, Texas 78712-1081, USA} \author{V. Khudik}
 \affiliation{Department of Physics, The University of Texas at Austin, Austin, Texas 78712-1081, USA} \affiliation{Institute for Fusion Studies, The University of Texas at Austin, Austin, Texas 78712, USA}\author{ G. Shvets}
 \affiliation{Department of Physics, The University of Texas at Austin, Austin, Texas 78712-1081, USA} \affiliation{Institute for Fusion Studies, The University of Texas at Austin, Austin, Texas 78712, USA}\author{M. C. Downer}
  \email{downer@physics.utexas.edu}
  \affiliation{Department of Physics, The University of Texas at Austin, Austin, Texas 78712-1081, USA}
\date{\today}% It is always \today, today,
             %  but any date may be explicitly specified

\date{\today}

\begin{abstract}

We present an in-depth experimental-computational study of the parameters necessary to optimize a tunable, quasi-monoenergetic, efficient, low-background Compton backscattering (CBS) x-ray source that is based on the self-aligned combination of a laser-plasma accelerator (LPA) and a plasma mirror (PM). The main findings are:  (1) an LPA driven in the blowout regime by 30 TW, 30 fs laser pulses producesnot only a high-quality, tunable, quasi-monoenergetic electron beam, but also a high-quality, relativistically intense ($a_0 \sim 1$) spent drive pulse that remains stable in profile and intensity over the LPA tuning range.  (2) A thin plastic film near the gas jet exit retro-reflects the spent drive pulse efficiently into oncoming electrons to produce CBS x-rays without detectable bremsstrahlung background.  Meanwhile anomalous far-field divergence of the retro-reflected light demonstrates relativistic ``denting" of the PM.  Exploiting these optimized LPA and PM conditions, we demonstrate \textit{quasi-monoenergetic} (50\% FWHM energy spread), \emph{tunable} (75 to 200 KeV) CBS x-rays, characteristics previously achieved only on more powerful laser systems by CBS of a split-off, counter-propagating pulse.   Moreover, laser-to-x-ray photon conversion efficiency ($\sim6 \times 10^{-12}$) exceeds that of any previous LPA-based quasi-monoenergetic Compton source.  Particle-in-cell simulations agree well with the measurements.

\end{abstract}

%Uncomment for PACS numbers title message
%\pacs{00.00, 20.00, 42.10}
% Keywords required only for MST, PB, PMB, PM, JOA, JOB? 
%\vspace{2pc}
%\noindent{\it Keywords}: Article preparation, IOP journals
% Uncomment for Submitted to journal title message
%\submitto{\JPA}
% Comment out if separate title page not required
\maketitle

%%%%%%%%           SECTION I                %%%%%%%%%%%%%%%%%%%%%%%%%%%%%%

\section{INTRODUCTION}

Generation of highly directional, narrow bandwidth hard x-ray or $\gamma $-ray beams by Compton backscatter (CBS) from relativistic electron beams \cite{Catravas01} has many applications including radiation therapy \cite{Weeks97}, radio surgery \cite{Girolami96}, industrial CT scanning \cite{Ketcham01}, homeland security \cite{Jannson07}, and photo-nuclear spectroscopy \cite{Schreiber2000, Kwan11}.  For such applications, narrow bandwidth CBS x-rays offer higher signal-to-noise ratio than broadband bremsstrahlung x-rays.  High quality CBS x-rays were demonstrated more than a decade ago using conventional electron accelerators \cite{Schonlein96,Leemans96,Gibson04}. Within the past decade, however, tabletop laser-plasma accelerators (LPAs) \cite{Tajima79, Esarey09} that accelerate electrons quasi-monoenergetically \cite{Mangles04, Geddes04, Faure04} to hundreds of MeV \cite{Hafz08, Kneip09, Froula09} or GeV \cite{Wang13, Kim13,Leemans14} energy within millimeters to centimeters
%, rather than hundreds of meters as in conventional accelerators, 
have emerged, opening the possibility of compact CBS x-ray sources compatible with small university laboratories \cite{Corde13, Hartemann07, Korz11}.

CBS x-rays have been generated from LPAs by two methods. In the first, a 100 TW laser system supplied both $1.9$~J, 35~fs LPA drive pulses and $0.5$~J, 90 fs split-off backscatter pulses. The latter were focused to spot size $w_0 = 22 \mu$m, intensity $a_0 \approx 0.3$ onto electrons emerging from the LPA to generate quasi-monoenergetic tunable CBS x-rays up to MeV energy \cite{Chen13,Power14}.   Here $a_0 \equiv eE_{\rm L}/m\omega c = 0.85 \sqrt{\lambda^2(\mu {\rm m})I(10^{18} {\rm W/cm}^2)}$ --- where $E_{\rm L}$ is the electric field of the laser pulse of frequency $\omega$, wavelength $\lambda$, intensity $I$, and $e$ and $m$ are electron charge and mass, respectively --- is a dimensionless laser strength parameter defined such that laser-electron interactions are relativistic for $a_0 \agt 1$. Under these conditions, stable overlap of backscatter pulse and LPA electrons was achieved despite shot-to-shot pointing fluctuations, as indicated by the high reproducibility ($>93\%$) and photon number stability ($60\%$) of the CBS x-rays \cite{Chen13,Power14}. In the second method, a 30 TW laser system directly supplied only $\sim 1$~J, $35$~fs LPA drive pulses. A plasma mirror (PM) then retro-reflected the drive pulse into the trailing relativistic electrons after the LPA \cite{Phuoc12}.  This method is self-aligning, and thus eliminates sensitivity to laser pointing fluctuations even for very small spot sizes.  It is thus a potentially attractive LPA-based Compton x-ray source for laboratories with smaller (tens of TW) laser systems.  However, \emph{tunable}, \emph{quasi-monoenergetic} x-rays have not yet been demonstrated by this method, only broadband x-rays centered at $\sim 50$ KeV \cite{Phuoc12}.  Moreover, key parameters that determine x-ray brightness, such as the intensity and spatial profile of the laser pulse after driving the LPA and reflecting from the PM, have not been measured, let alone optimized. 

Here we report an in-depth study of the parameters necessary to optimize CBS x-ray generation --- \textit{i.e.} to generate \emph{tunable}, \emph{quasi-monoenergetic} CBS x-rays with \emph{high conversion efficiency} and \emph{low background} --- using the self-aligned combination of LPA and PM (the second method).  In this study, we fully characterize the laser pulse spatial intensity profile immediately after driving the LPA and after reflecting from the PM by both measurement and simulation. There are three major findings from the study.  First, a mildly relativistic ($a_0 \approx 1.6$) incident laser pulse remains relativistic ($1 \alt a_0 \alt 2$) and of high beam quality after driving an LPA in the bubble regime.  Moreover, its intensity profile remains stable as the LPA plasma density $\bar{n}_e$ changes from $1.4$ to $2.2 \times 10^{19}$ cm$^{-3}$, a range over which electrons remain quasi-monoenergetic ($10-20\%$ FWHM energy spread) and collimated, but tune in energy from 60 to 90 MeV.  This $a_0$ exceeds that achieved using a split-off backscatter pulse, suggesting that a future nonlinear Compton source \cite{NLCBS,NLCBSEXP} may be more readily achievable for some LPAs via the PM method.  
Second, the PM yielded near unity reflectance at $a_0 \sim 1$, an intensity regime for which PM reflectivity has not been well characterized.  This suggests that prepulses, a major source of declining PM reflectivity at relativistic intensity, are suppressed by the act of driving an LPA in the bubble regime. Moreover, use of a plastic film only 90 $\mu$m thick for the PM rendered transmitted bremsstrahlung radiation undetectable, resulting in a very high signal-to-noise ratio Compton source.
Third, our measurements of far-field angular divergence of the retro-reflected drive pulse together with simulations of its interaction with the PM show that the PM surface curved relativistically\cite{Dromey09,Vincenti14}, confirming that the spent drive pulse was relativistically intense, and suggesting that higher x-ray yield may be achievable in future work by optimizing this curvature to focus the retro-reflected drive pulse onto trailing electrons.

Based on findings of this study, we demonstrate generation of quasi-monoenergetic ($50\%$ FWHM energy spread), tunable (75 to 200 keV photon energy) CBS x-rays by the LPA-PM method for the first time. X-ray energy was tuned by varying $\bar{n}_e$, and thus electron energy, over a range that preserved narrow electron energy spread and stable, relativistic spent drive pulse intensity profile.  Previously, quasi-monoenergetic, tunable CBS x-rays had been generated from LPAs only by the split-off pulse method\cite{Chen13,Power14}.  Moreover, we demonstrate photon conversion efficiency $\sim 6 \times 10^{-12}$ from laser pulse to x-rays that is higher than achieved so far using a split-off scattering pulse by factors ranging from six \cite{Chen13} to thirty \cite{Power14}.  It is thus the highest conversion efficiency so far demonstrated for a LPA-based quasi-monoenergetic Compton source. This high efficiency is the result of our coordinated achievement, and detailed characterization, of a stable spent LPA drive pulse profile with $a_0 \agt 1$ that reflects with near-unity efficiency from a PM that guarantees excellent overlap with $150$ pC, quasi-monoenergetic electron bunches.  This unique convergence of conditions has not been achieved, nor characterized, in previous work.  This, of course, does not rule out the possibility that higher conversion efficiency could be achieved by improving overlap parameters in the split-off pulse method, or by exploiting relativistic PM curvature more effectively in the PM method.  Higher conversion efficiency has been achieved by operating the LPA in a regime that produces broadband electrons\cite{Phuoc12}, but at the cost of $100\%$ x-ray energy spread and loss of tunability.

%%%%%%%%%%
%     FIGURE 1     %
%%%%%%%%%%

\begin{figure*}[htb]
\centering
\includegraphics[width=0.9\textwidth]{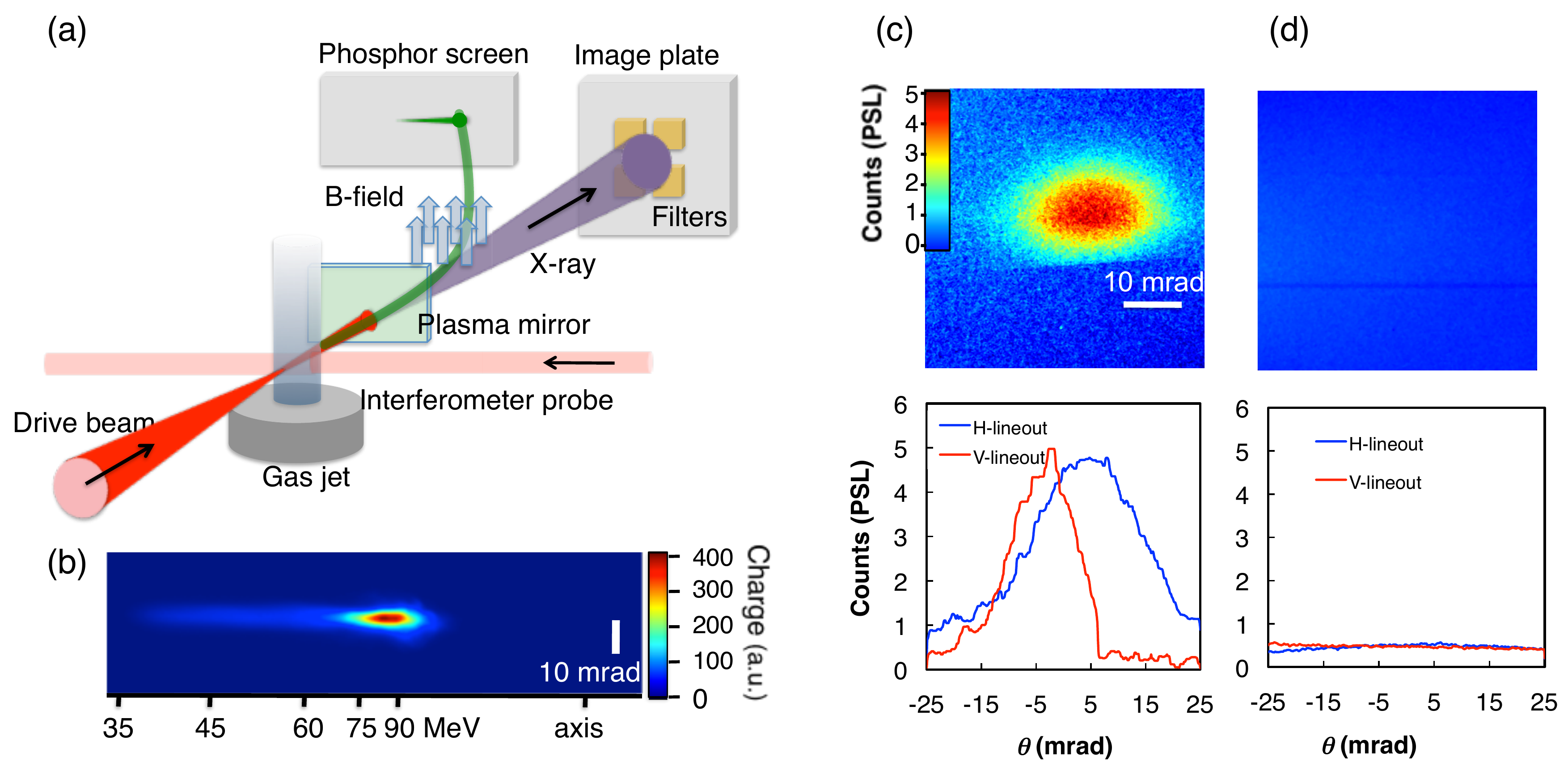}
\caption{\textbf{(a)} Schematic experimental setup for Compton backscatter (CBS).  Interaction of a drive laser (red) with a plasma created within the plume of a gas jet accelerates electrons (green), which a magnetic field deflects onto a phosphor screen. The laser pulse ionizes a thin plastic foil placed at the exit of the plasma accelerator, forming a plasma mirror (PM). The retro-reflected drive laser pulse backscatters from the electron beam after the accelerator, creating an x-ray beam (purple) that an imaging plate records after the x-rays pass through a filter pack. \textbf{(b)} phosphor screen image of the electron spectrum at plasma density $2.2\times10^{19}\rm\ cm^{-3}$, showing peak at 90 MeV. \textbf{(c)} and \textbf{(d)} x-ray beam profile (top), and horizontal and vertical lineouts (bottom), accumulated over 10 shots with $90 \mu$m-thick plastic PM placed either \textbf{(c)} at gas jet exit, yielding strong CBS x-rays, or \textbf{(d)} 15 mm downstream from gas jet exit, showing no detectable bremsstrahlung background. }
\label{figure_setup}
\end{figure*}

%%%%%%%%           SECTION II           %%%%%%%%%%%%%%%%%%%%%%%%%%%%%%

\section{EXPERIMENTAL PROCEDURE}

Experiments utilized the 30-TW UT$^{3}$ laser system at the University of Texas at Austin, which operates with 10-Hz repetition rate at central wavelength 800 nm (photon energy $E_L = 1.5$ eV). Fig. 1(a) shows the schematic setup. To drive the LPA, $30$ fs, $0.8$ J linearly-polarized laser pulses were focused with $f$-number 12.5 to Gaussian spot radius $w_0 = 10\pm 1 \mu$m (intensity profile FWHM = $w_0\sqrt{2\ln 2} = 11.8\pm 1 \mu$m), where the uncertainties indicate rms shot-to-shot fluctuations, and peak intensity $6\times 10^{18} \rm\ W/cm^2$ ($a_0 \approx 1.6$) at the entrance of a 1-mm supersonic gas jet (99\% helium, 1\% nitrogen mixture).  A transverse interferometer measured time-averaged plasma density profile $\bar{n}_e(r,z)$ on each shot.  A magnetic electron spectrometer placed downstream from the gas jet analyzed electron energy on each shot.  It consisted of a 1 T magnet that deflected electrons onto a terbium-activated gadolinium oxysulfide (Gd$_2$O$_2$S:Tb) phosphor screen (Kasei Optonix model Kyokko PI200) from which electron-induced fluorescence at $545$ nm was imaged onto a 12-bit charge-coupled device (CCD) camera.  The divergence, energy and charge of the electron beam were optimized by controlling $\bar{n}_e$, focal spot longitudinal position and drive pulse duration \cite{Tsai12}. Electrons were produced in the ionization-injected \cite{Pak10,McGuffey10} bubble regime \cite{Pukhov02} with peak energy as high as $\sim 90$ MeV [Fig. 1(b)], with energy spread (FWHM) $10\%$ under optimum conditions, and no larger than $20\%$ for all data presented here, $\sim 4$ mrad divergence, and integrated charge $\sim 150$ pC for energies $> 30$ MeV.  Energy spread was determined by multiplying the vertically-integrated, magnetically-dispersed recorded electron trace $dN_e/dx$ by the magnet dispersion $dx/dE$ to generate an energy distribution $dN_e/dE$, then taking the ratio of the width $E_{\rm FWHM}$ of the quasi-monoenergetic peak to its central energy $E_{\rm peak}$.  Here $N_e$ is electron number, $E$ electron energy and $x$ horizontal distance along the phosphor screen. The cited energy spread takes into account the low-energy tail of the electron distribution. Charge was determined from integrated fluorescent photon number emitted from the phosphor screen using published calibrations for PI200 \cite{Wu12}.  Electron energy was tuned continuously from $\sim 60$ to $90$ MeV while retaining narrow energy spread and collimation by tuning gas jet backing pressure to vary $\bar{n}_e$ from $1.7$ to $2.2\times10^{19}\rm\ cm^{-3}$.

Conversion efficiency from laser pulse to CBS x-rays depends critically on the intensity profile of the spent drive pulse transmitted through the LPA.  We therefore measured transverse laser intensity profile at the LPA exit, with the PM temporarily removed, by reflecting the spent drive pulse into an f/10 relay imaging system with a 2-inch diameter pellicle inserted 15 cm downstream of the gas jet.   The pellicle left the electron beam unperturbed, so the exit spot profile measurement could be correlated with electron beam properties.  The absolute transmitted intensity was then estimated from the measured profile and transmitted energy, using an estimated pulse duration of 30 fs. We validated this estimate by simulating the transmitted pulse properties for our experimental conditions, as discussed further in Sec. IV. 
%had a spot size of  $\sim$40 $\mu$m (FWHM) with $\sim50\%$ of its original energy as shown in Fig.5(b), corresponding to a peak intensity $1.1\times 10^{18} \rm\ Wcm^{-2}$ ($a_0 = 0.7$), 

Conversion efficiency also depends critically on PM reflectivity.  PMs are widely used at \emph{sub}-relativistic incident intensities ($a_0 < 1$) to improve the temporal contrast of ultrashort laser pulses that are subsequently focused to ultra-relativistic intensity \cite{Kapteyn91,Ziener03,Doumy04,Geissel11}. Here the intensity incident on the PM is mildly relativistic ($a_0 \agt 1$), a range in which PM reflectivity is not well characterized and can depend sensitively on pulse contrast and intensity.  We therefore directly measured PM performance at relativistic intensity \emph{without} the gas jet using well-characterized, high-contrast 30 fs pulses incident at $5^\circ$ from the normal with peak fields in the range $0.2 \alt a_0 \alt 1.6$.  Pulses with $a_0 \agt 1$ can depress the PM surface due to their high ponderomotive pressure\cite{Dromey09, Vincenti14}, causing the reflected pulse to focus in front of the PM surface and diverge far from it.  This relativistic focusing could potentially increase the intensity that interacts with the trailing electrons, and thereby improve conversion efficiency.   We therefore characterized far-field divergence of the time-integrated reflected light using two different collection cones:  (\textit{i}) a wide ($f/5$) cone, consisting of an energy calorimeter with 6 cm aperture placed 30 cm from the PM surface; (\textit{ii}) a narrow ($f/20$) cone, in which reflected light was relay imaged from the PM surface to a CCD camera.  Method (\textit{i}) determined spatially-integrated reflectivity from the ratio of the measured energy to the incident laser pulse energy.  Method (\textit{ii}) determined reflectivity with $\sim 10\mu$m spatial resolution.   In this case, reflectivity was determined from the ratio of peak intensity of the reflected beam to the peak incident intensity.
For CBS x-ray generation, the PM was inserted $\sim500\mu$m from the jet exit plane, and reflected the spent pulse at $175\,^{\circ}$ to the e-beam direction, to avoid direct retro-reflection into the laser amplifier system.  PMs made of 90-$\mu$m plastic (household cellophane), 1-mm thick plastic and 1-mm thick fused silica (microscope slide) were tested; all perturbed the transmitted electrons properties only slightly.  The PM was translated 500$\mu$m transversely after each shot to remove the damaged spot from the path of the next laser pulse, and could be translated longitudinally as far as 2 cm from the jet.  Since CBS efficiency fell rapidly to undetectable levels as PM distance from the jet increased, due to expansion of the spent drive pulse, background bremsstrahlung x-rays generated by the highly collimated electrons inside the PM could be characterized at large separations. 
%$>90\%$ was retro-reflected by a PM 
% However, the 90-$\mu$m thick plastic foil can achieve the same reflectivity to produce CBS x-rays without detecting bremsstrahlung background.
%Electron beams are generated, and measured to be peaked at $\sim100$ MeV at the optimal plasma density at $2.0\times10^{19}\rm\ cm^{-3}$, with  and a lanex screen imaged onto a charge coupled device (CCD) camera.

%%%%%%%%%%
%     FIGURE 2     %
%%%%%%%%%%

\begin{figure*}[t]
\centering
\includegraphics[width=1\textwidth]{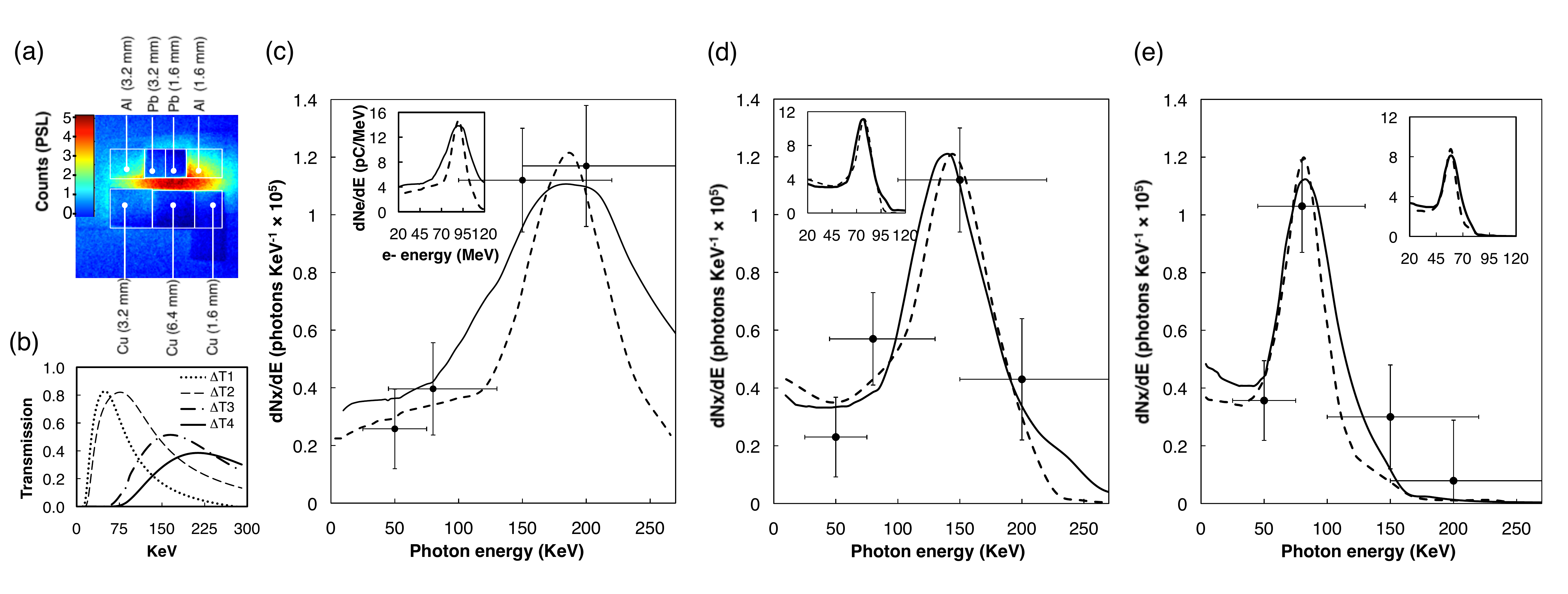}
\caption{CBS x-ray spectra. \textbf{(a)} CBS x-ray beam profile accumulated over 10 shots through Al, Cu, and Pb filter combinations.   \textbf{(b)} Plot of transmission difference ($\Delta T$) spectra of four filter pairs: 1) Al (1.6 mm) and Cu (1.6 mm);  2) Al(3.2 mm) and Cu(3.2 mm); 3) Pb (1.6 mm) and Cu (6.4 mm); 4) Cu (6.4 mm) and Pb (3.2 mm), providing 4 energy bandpass filters peaked at 50, 80, 150 and 200 KeV with FWHM bandwidth of 50, 70, 100, and 120 KeV, respectively. \textbf{(c)-(e)} CBS x-ray spectra for 3 values of LPA plasma density $\bar{n}_e$ and electron energy $E$:  \textbf{(c)} $\bar{n}_e = 2.2 \times 10^{19}$ cm$^{-3}$, $E = 90$ MeV; \textbf{(d)} $\bar{n}_e = 1.8 \times 10^{19}$ cm$^{-3}$, $E = 75$ MeV;  \textbf{(e)} $\bar{n}_e = 1.4 \times 10^{19}$ cm$^{-3}$, $E = 60$ MeV.  Data points:  mean energy measured in each of the 4 bands. Horizontal error bars:  FWHM of each band.  Vertical error bars:  RMS variation for multiple 10-shot data sets. Solid curves:  10-shot-averaged x-ray spectra calculated from electron spectra for each of 10 shots. Dashed curves:  \emph{single-shot} x-ray spectra calculated from single-shot electron spectra.   Inset of panel \textbf{(c)-(e)}:  measured \emph{electron} spectrum at $\bar{n}_e = 2.2, 1.8, and 1.4 \times 10^{19}$ cm$^{-3}$ averaged over 10 shots (solid curve), and single-shot (dashed curve), corresponding to the x-ray spectrum shown in main panel. The axis labels are the same.}
 
\label{figure_xspectr}
\end{figure*}

CBS yields x-ray photons of energy $E_X =4\gamma^{2} E_{\rm L}$, where $\gamma$ is the electron Lorentz factor.  Thus for $E_{\rm L} = 1.5$ eV, and 90 MeV ($\gamma = 180$) electrons, 185 KeV x-ray photons were expected.  A $50 \times 50$-mm imaging plate (IP, Fujifilm BAS-IP MS 2025 E) placed inside the vacuum chamber 0.8 m from the scattering point detected these photons with high spatial resolution over $\sim$ 60 mrad divergence angle. Because of the low detection efficiency ($\sim0.7$ mPSL/photon \cite{IP08}, where PSL denotes photo-stimulated luminescence), x-ray data was accumulated over $\sim 10$ shots to achieve adequate signal-to-noise ratio for quantitative analysis.  We measured the spatially-averaged x-ray spectrum by comparing transmission through a set of four filter pairs, each composed of different elements, placed symmetrically in horizontal rows immediately (within 5 mrad) above and below the horizontal center line of the x-ray profile [Fig. 2(a)], in front of the IP \cite{Chen13,Phuoc12}:  1) Al (1.6 mm) and Cu (1.6 mm); 2) Al (3.2 mm) and Cu (3.2 mm); 3) Pb (1.6 mm) and Cu (6.4 mm); 4) Pb (3.2 mm) and Cu (6.4 mm).  The differential transmission spectra $\Delta T_i(h\nu_X)$, $i = 1-4$) of these filter pairs, plotted in Fig.~\ref{figure_xspectr}(b), provided energy bandpass filters peaked at 50, 80, 150 and 200 KeV with FWHM bandwidths 50, 70, 100, and 120 KeV, respectively.  The transmitted signal through each pair along with IP response curve \cite{IP08} was used to determine the number of photons with energies within the filter pair's transmission band.  To average the spectra spatially, and avoid systematic errors due to inaccuracies or non-uniformities in filter thickness, actuators translated the filter pack horizontally across the beam profile between each of several 10-shot sequences.  In addition, data was recorded with filters in different horizontal orders.  Within error bars, these variations did not change the extracted spectra.  Apart from these filters, the x-ray beam passed through only the $90 \mu$m-thick PM and a downstream $20 \mu$m-thick Al film (not shown in Fig.~1) that deflected laser light transmitted through the PM to a beam dump.  These affected the x-ray transmittance and spectrum negligibly.  X-ray photon number was obtained by integrating net PSL counts over the x-ray profile recorded on the IP, taking into account the measured x-ray spectrum and the IP response curve \cite{IP08}. 
%A gate valve was between the IP extension tube and the chamber, so it can be accessed without breaking the chamber vacuum. 

%%%%%%%%%%
%     FIGURE 3     %
%%%%%%%%%%
\begin{figure*}[t]
\centering
\includegraphics[width=0.9\textwidth]{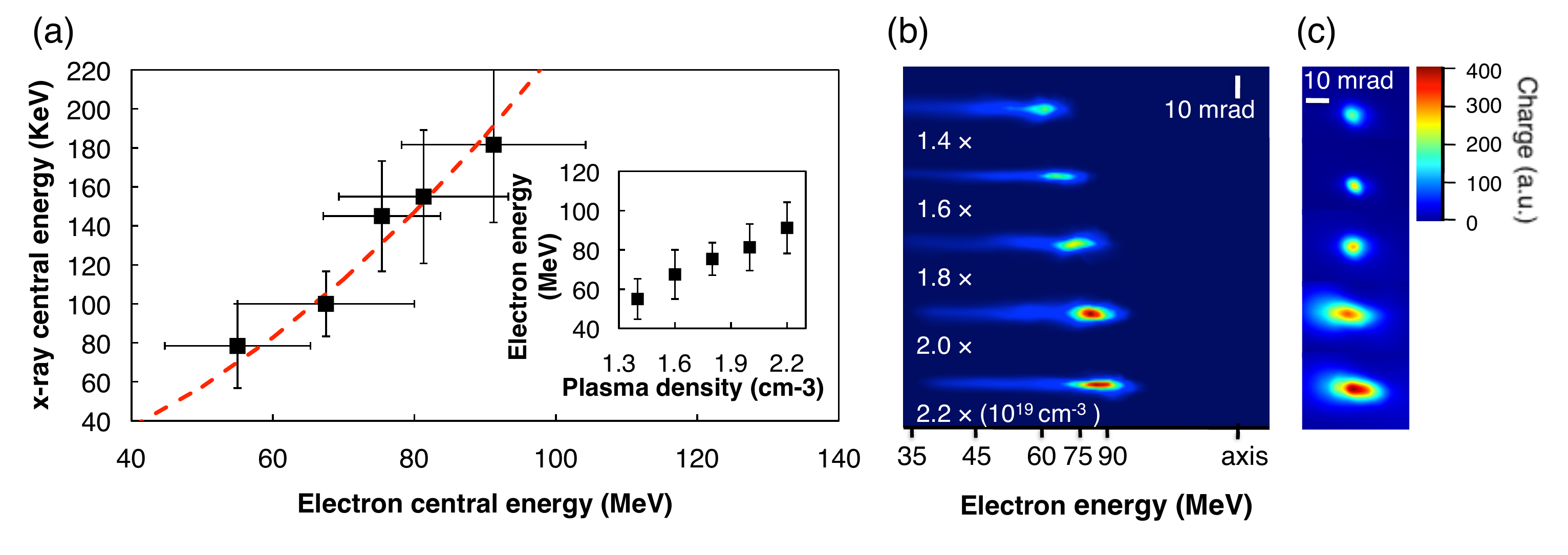}
\caption{\textbf{(a)} Measured x-ray central energy, plotted vs. measured electron central energy, compared to 4$\gamma^{2}$ scaling (red dashed line). Inset:  electron energy vs. plasma density of a 1-mm gas jet. Each point is the average of 10 shots taken with the same gas-jet position and backing pressure. Error bars represent FWHM of 10-shot averaged e-beam bandwidth. \textbf{(b)} Electron spectra observed on PI200 phosphor at five plasma densities: 1.4, 1.6, 1.8, 2.0, and 2.2 $\times 10^{19} cm^{-3}$.  \textbf{(c)} Electron beam profiles recorded 15 cm downstream of gas jet on PI200 phosphor screen with magnet removed, at the same five densities. }
\label{figure_tune}
\end{figure*}

%%%%%%%%%%%%%%%%%%%%%%%%%%%%%%%%%%%%%%%%%%%%%%%%%%%%%%%%%%%%%%%%
%%%%%%%%%%%%%%%                SECTION III          %%%%%%%%%%%%%%%%%%%%%%%%%%%%%%%%%

\section{EXPERIMENT RESULTS}

%%%%%%%%%%%%%%%%%%%%      SUBSECTION III -1        %%%%%%%%%%%%%%%%%%%
\subsection{\label{sec:level2}CBS x-ray and e-beam properties}

Fig.~1(c) (upper panel) shows a background-subtracted 10-shot image of the x-ray beam, unobstructed by filters, obtained for the same conditions as the electron spectrum in Fig.~1(b). The x-ray beam diverged $\sim$20 (10) mrad (FWHM) in the horizontal (vertical) direction, as shown in the lower panel of Fig.~1(c).  Larger horizontal pointing fluctuations of the e-beam ($\sim10$ mrad) were primarily responsible for the asymmetry in the x-ray profile.  X-rays were observed for 95$\%$ of shots for which the LPA produced quasi-monoenergetic relativistic electrons and the PM was placed at the gas jet exit, although 10-shot averaging was necessary to achieve adequate signal-to-noise ratio to fit a Gaussian curve to the x-ray profile. On most of the 5\% of shots that failed to produce x-rays, we observed anomalous large-angle scatter or anomalously low reflectivity from the PM on a nearby detector, suggesting that the transmitted drive pulse struck a local defect on the plastic film. No x-rays were observed in any shots for which the LPA failed to produce relativistic electrons, or the gas jet was turned off.  When the PM was moved 15 mm or further away from gas jet, at most a background bremsstrahlung x-ray signal $\sim 10\times$ weaker than the signal shown in Fig.~1(c) was observed in conjunction with relativistic electrons.  With a 90-$\mu$m plastic foil PM, no background bremsstrahlung was detected, as shown in Fig. 1(d).  All results presented here were obtained with this PM.

 Fig.~2(a) shows a typical 10-shot-averaged x-ray profile transmitted through the filter pack. 
%The filters were arranged in the upper and lower half slightly away from center to allow accurate reconstruction of the unfiltered x-ray spatial profile and alignment of filters according to the beam center. 
Figs.~2(c)-(e) show extracted spatially-averaged x-ray spectra (black dots with error bars) corresponding to the electron spectra shown in the insets. The x-ray spectra are clearly \emph{not} decaying exponentially with increasing photon energy, as in previously reported PM/LPA-based CBS results \cite{Phuoc12}.  Instead spectral intensity rises below, and falls beyond, a peak x-ray energy.   
%A Gaussian curve fitted to the data points yields a spectrum centered at 180 KeV with energy spread $\delta E_X/E_X = 0.61$, where $\delta E_X$ is the FWHM of the fitted Gaussian profile.  
The dashed curves in main panels (c)-(e) are CBS spectra calculated from \emph{single-shot} electron spectra (dashed curves, insets) within the 10-shot sequence that produced each set of x-ray data.  The calculated peak energies [(c) 185 keV; (d) 144 keV; (e) 81 keV] agree well with the data points, but the calculated peak widths ($\delta E_X(FWHM)/E_X$) are narrower: (c) $0.44$; (d) 0.51; (e) 0.5, respectively.   This discrepancy is removed when observed shot-to-shot fluctuations of the electron spectra are taken into account, yielding the 10-shot averaged electron spectra shown by solid curves in the insets of Figs.~2(c)-(e).  The solid curves in the main panels of Figs.~2(c)-(e) show the corresponding calculated 10-shot averaged x-ray spectra, which agree very well with the data, and are somewhat broader:  (c) 0.63; (d) 0.56; (e) 0.66.  The good agreement with the 10-shot averaged x-ray data points validates the extracted \emph{single-shot} FWHM, which averages 0.5 over our tuning range.  

%\subsection{CBS x-ray energy tuning}

The central x-ray energy $E_X$ was tuned from $\sim$75 to $\sim$200 KeV [Fig.~3(a), main panel] by tuning the central energy of quasi-monoenergetic LPA electrons from 60 MeV ($\gamma = 120$) to 90 MeV ($\gamma = 180$) [Fig.~3(b)] by tuning plasma density from 1.4 to $2.2\times10^{19}$ cm$^{-3}$ in five increments [Fig.~3(a), inset].  The horizontal (vertical) error bars in the main panel (inset) of Fig.~3(a) represent the 10-shot-averaged e-beam energy bandwidth (FWHM). The vertical error bars in the main panel of Fig.~3(a) represent the corresponding FWHM of the peaked x-ray spectra. The tuned x-ray energy agrees well with the theoretical scaling $E_X =4\gamma^{2} E_L$, shown by the dashed red curve in Fig.~3(a).  

%%%%%%%%%%
%     FIGURE 4     %
%%%%%%%%%%
\begin{figure*}[t]
\centering
\includegraphics[width=0.8\textwidth]{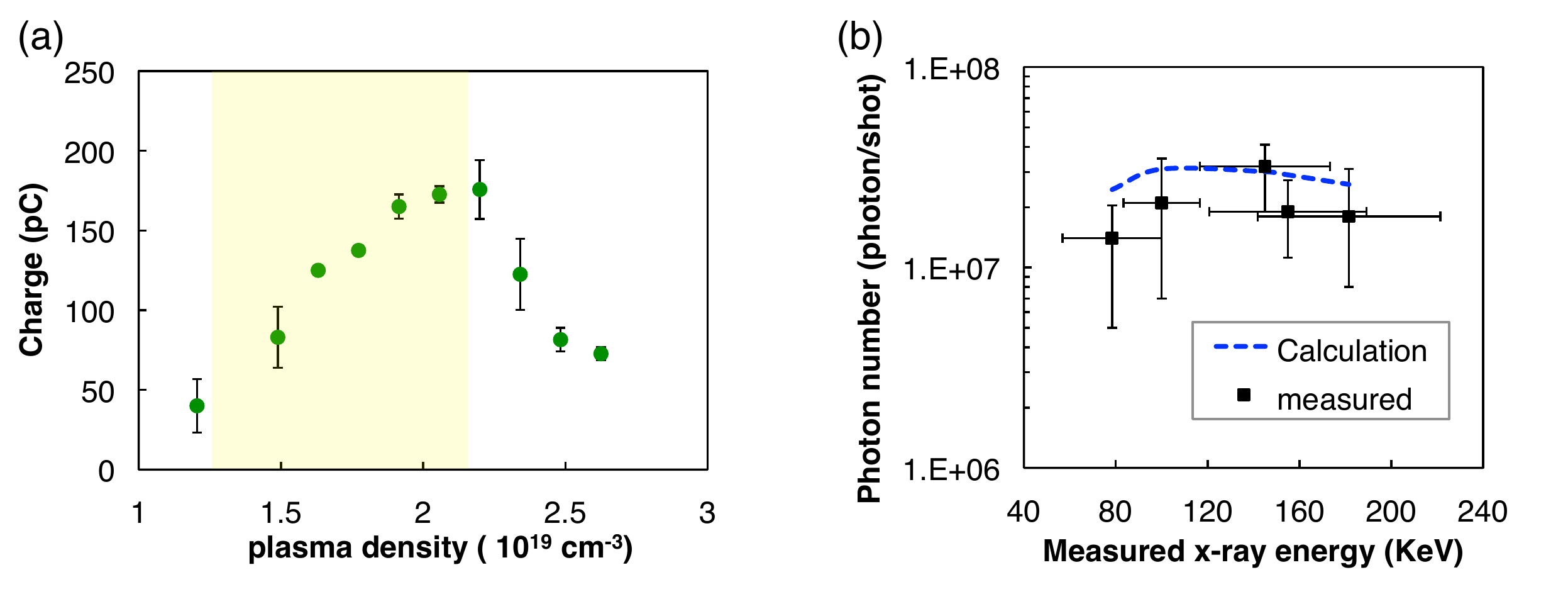}
\caption{\textbf{(a)} Measured electron charge from the LPA (green data points) as a function of plasma density $\bar{n}_e$.   Shaded yellow region denotes $\bar{n}_e$ range over which the LPA produces quasi-monoenergetic, central-energy-tunable electrons. 
\textbf{(b)}  Measured (black dots) and calculated (blue dashed curve) x-ray photon number per shot plotted vs. average central energy of the x-ray beam. Horizontal error bar represents the FWHM of the Gaussian curve fitted to the measured x-ray spectrum. Vertical error bar represents uncertainty caused by fluctuation of background noise relative to the signal.  }
\label{figure_xn_top}
\end{figure*}

Fig.~4(a) shows the total electron charge (green data points) from the LPA as a function of $\bar{n}_e$.  The yellow shaded region denotes the $\bar{n}_e$ range over which the LPA produced quasi-monoenergetic electrons, in which charge varied from 90 to 160 pC.  Fig.~4(b) shows the x-ray photon number per shot (data points with error bars) plotted as a function of central x-ray photon energy, obtained over the same $\bar{n}_e$ range.   Photon number was stable to within 50$\%$ of its average value 2 $\times 10^{7}$ throughout the tuning range.  The blue dashed curve is a calculated photon number, discussed in Sec. IV. 

%%%%%%%%%%
%     FIGURE 5     %
%%%%%%%%%%
\begin{figure}[htp]
\centering
\includegraphics[width=0.41\textwidth]{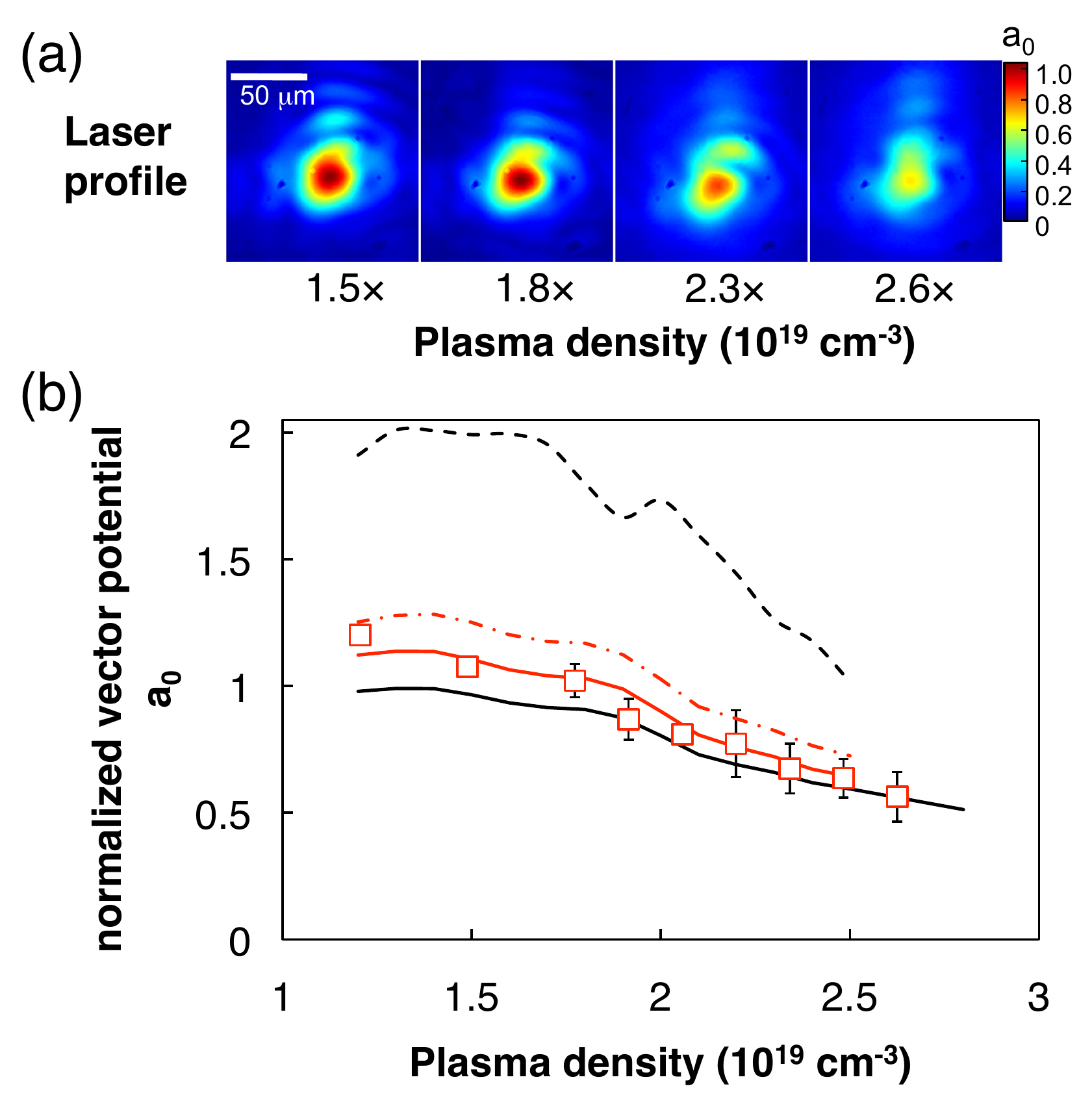}
\caption{\textbf{(a)} Measured laser beam profiles at exit plane of LPA, from which measured $a_0$ values in panel (b) were derived.  \textbf{(b)} Measured (red squares) and simulated (curves) on-axis, time-averaged $a_0$ of the laser pulse at the gas jet exit as a function of plasma density $\bar{n}_e$.   Black dashed curve: simulated $a_0$ assuming a sharply bounded plasma exit profile, with no density down-ramp. Red dot-dashed curve:   simulated $a_0$ assuming a 200 $\mu$m down-ramp similar to the measured $\bar{n}_e(z)$ exit profile. Red solid curve:  $a_0$ obtained by averaging the calculated $a_0$ profiles of the previous simulation over a transverse area of 6 $\mu$m diameter, to mimic spatial resolution of the detector.  Black solid curve:  calculated $a_0$ of the scattering pulse after reflecting from PM, obtained by multiplying $a_0$ from red dot-dashed curve by PM reflectivity (Fig.~6, solid blue curve). }
\label{figure_a0}
\end{figure}

%%%%%%%%%%%%%%%%%%%%      SUBSECTION III -2     %%%%%%%%%%%%%%%%%%%
\subsection{Laser intensity after LPA and PM}

Fig.~5(a) shows images of drive pulse $a_0$ profiles at the gas jet exit for four $\bar{n}_e$.  Although the shapes of these profiles change little with density, peak $a_0$ decreased about $40\%$  as $\bar{n}_e$ increased from 1.7 to $2.6 \times 10^{19}$ cm$^{-3}$. The red squares in Figure~5(b) show how the peak field strength $a_0$ of the drive pulse at the exit plane of the gas jet varied as a function of $\bar{n}_e$.  The absolute $a_0$ scale was estimated by assuming all measured transmitted pulse energy was contained within the imaged exit profile and that the pulse maintained $30$ fs duration.  Since the latter are crude approximations, the absolute $a_0$ scale was further validated through PIC simulations, as discussed in Sec. IV.  

The top row of Fig.~3(c) shows $\bar{n}_e$-dependent profiles of electrons recorded on a PI200 screen 35 cm downstream from the LPA with the magnet removed.  Throughout the range $1.4 < \bar{n}_e < 2.2 \times 10^{19}$ cm$^{-3}$ in which the LPA produced quasi-monoenergetic electrons, the electron beam diverges less than $10$ mrad, and thus expands negligibly within the $500 \mu$m distance between gas jet exit and PM.  Since the electron beam exits the LPA typically with only a few microns diameter \cite{Chen13}, it is much narrower than the FWHM of the transmitted drive pulse at the point of backscatter, and can be assumed to interact with its peak intensity.  For $\bar{n}_e > 2.2\times10^{19}$ cm$^{-3}$, the electron beam divergence increases rapidly to $>20$ mrad (not shown).

%\subsection{Plasma mirror reflectivity at relativistic laser intensities}
 
%As explained in the previous Sections, the plasma mirror in our experiment reflects the LWFA driver and achieve a counter-propagating setup to collide with an accelerated electron bunch. Plasma mirrors are routinely used at sub-relativistic intensities (below $10^{18}$ W/cm$^{2}$) to improve the temporal contrast of an ultrashort laser pulse that is subsequently focused to an ultra-relativistic intensity \cite{Kapteyn91,Ziener03,Doumy04,Geissel11}. The beam intensity in our case is outside of the conventional range since we use a plasma mirror to reflect the driver beam whose intensity can be as high as $5 \times 10^{18}$ W/cm$^{2}$.

%%%%%%%%%%
%     FIGURE 6      %
%%%%%%%%%%
\begin{figure}[htb]
\centering
\includegraphics[width=0.5\textwidth]{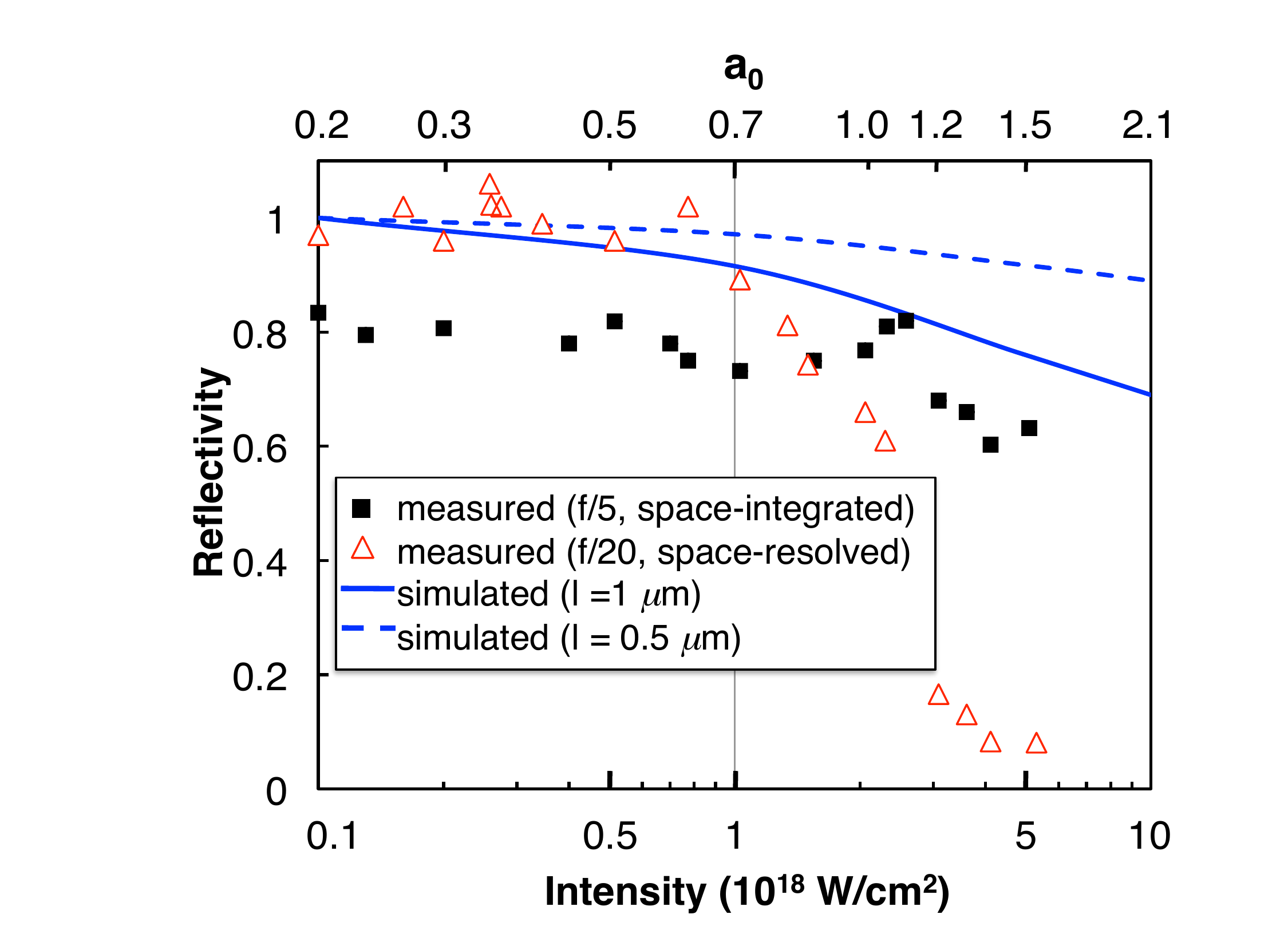}
\caption{Reflectivity of PM vs. intensity.  Red triangles:  measured reflectivity at peak of laser profile; black squares:  measured space- and time-integrated reflectivities, measured by energy meter;  blue curves:  simulated space- and time- integrated reflectivity, assuming a preformed plasma layer of density scale length of 1$ \mu$m (solid) and 0.5 $ \mu$m (dashed).}
\label{figure_ref_sim}
\end{figure}

Fig.~6 shows results of PM reflectivity measurements.   The spatially-integrated, wide ($f/5$) cone calorimeter measurements (black squares) yielded $\sim80\%$ reflectivity for incident intensities from $10^{17}$ to $2\times10^{18}$ W/cm$^{2}$, then dropped gradually to $\sim 60\%$ as intensity increased to $5 \times 10^{18}$ W/cm$^{2}$.   In contrast, the spatially-resolved, narrow ($f/20$) cone intensity measurements (red triangles) yielded nearly 100\% reflectivity at the peak of the profile for incident intensities from $10^{17}$ to $\sim 10^{18}$ W/cm$^{2}$, then dropped steeply to $< 10\%$ at $5 \times 10^{18}$ W/cm$^{2}$.   Evidently the spatially-integrated measurement yields only $80\%$ reflectivity for $I < 10^{18}$ W/cm$^2$ because it measures not only the intense center of the focal spot, for which reflectivity is near unity, but the less intense wings, for which reflectivity is only a few percent because overdense plasma is not created.  The result is straightforwardly explained if $20\%$ of the pulse energy lies outside the central focus and fails to reach a threshold intensity $\sim10^{17}$ W/cm$^{2}$, which previous work has shown is needed to create a highly reflective overdense plasma \cite{Ziener03}.  For relativistic intensity, on the other hand, the spatially-integrated measurement yields much \emph{higher} reflectivity.   This discrepancy could be explained either by strong absorption that is localized in the intense center of the pulse profile, or by defocusing of the reflected pulse outside the $f/20$ collection cone due to curvature of the PM surface induced by strong ponderomotive pressure.  Below we explore these mechanisms through simulations.  

%%%%%%%%%%%%%%%%%%%%%%%%%%%%%%%%%%%%%%%%%%%%%%%%%%%%%%%%%%%%%%
%%%%%%%%%%%%%%%%   SECTION IV           %%%%%%%%%%%%%%%%%%%%%%%%%%%%%%%%%%%

\section{DISCUSSION}
To help understand and calibrate the exit-plane intensity results in Fig.~5, we used the 3D particle-in-cell (PIC) code Virtual Laser Plasma Lab (VLPL) \cite {Pukhov99} to simulate a 30 fs (FWHM) Gaussian pulse with spot size $11\mu$m (FWHM) and peak intensity $5 \times 10^{18}$ W/cm$^{2}$ propagating into a 200 $\mu$m density up-ramp followed by an 800 $\mu$m plateau of constant $\bar{n}_e$.   For $1.4 < \bar{n}_e < 2.2\times10^{19}$ cm$^{-3}$, the simulated laser pulse self-focused, and formed a plasma bubble that trapped ionization-injected electrons before accelerating them to 60 to 90 MeV, in good agreement with the electron energy measurements in Fig.~3(b).  Elsewhere we have shown for similar conditions that the VLPL results agree well with bubble dynamics measured with an all-optical streak camera and with electron energy measurements \cite{Li14}.  Here we are concerned with laser intensity transmitted through the LPA.  The black dashed curve in Fig.~5(b) shows simulated time-averaged on-axis $a_0$ at the jet exit as a function of $\bar{n}_e$, for an artificially sharply-bounded jet exit with no density down-ramp.  The shape of the curve resembles the data (red squares), but is roughly $2\times$ higher than the crudely estimated absolute $a_0$ values of the data points.  The red dot-dashed curve shows the simulated on-axis exit-plane intensity when a realistic 200 $\mu$m density down-ramp, matching the profile observed in the transverse interferometer, was added after the 800 $\mu$m plateau. The intensity is now lower because the exiting pulse expands in the down-ramp.  The red solid curve shows the result of locally averaging the exit-plane intensity profile within a transverse area of 6 $\mu$m diameter, to mimic the spatial resolution of the relay imaging system.  This spatial resolution was calibrated directly by imaging a resolution test chart (1951 USAF) placed at the exit plane of the LPA.  The result agrees almost perfectly with the data points. Thus the crude procedure described earlier for estimating absolute $a_0$ values at the jet exit fortuitously agrees with 3D PIC simulations, which also corroborate the observed $\bar{n}_e$-dependence.   Most likely the assumption that the central imaged laser profiles [shown in Fig.~5(a)] contained \emph{all} of the transmitted pulse energy overestimated the actual energy in this profile, while the assumed 30 fs pulse duration overestimated the actual duration by neglecting front end erosion by the laser-plasma interaction.  These two errors evidently canceled, yielding agreement.  The dot-dashed red curve in Fig.~5(b) thus accurately represents exit-plane axial $a_0$ without detector averaging, which falls in the range $1.2 > a_0 > 0.9$ (peak intensity $2 > I > 1\times10^{18}$ W/cm$^2$) for the $\bar{n}_e$ range of interest.  

%%%%%%%%%%
%     FIGURE 7     %
%%%%%%%%%%
 \begin{figure*}[htb]
\centering
\includegraphics[width=0.8\textwidth]{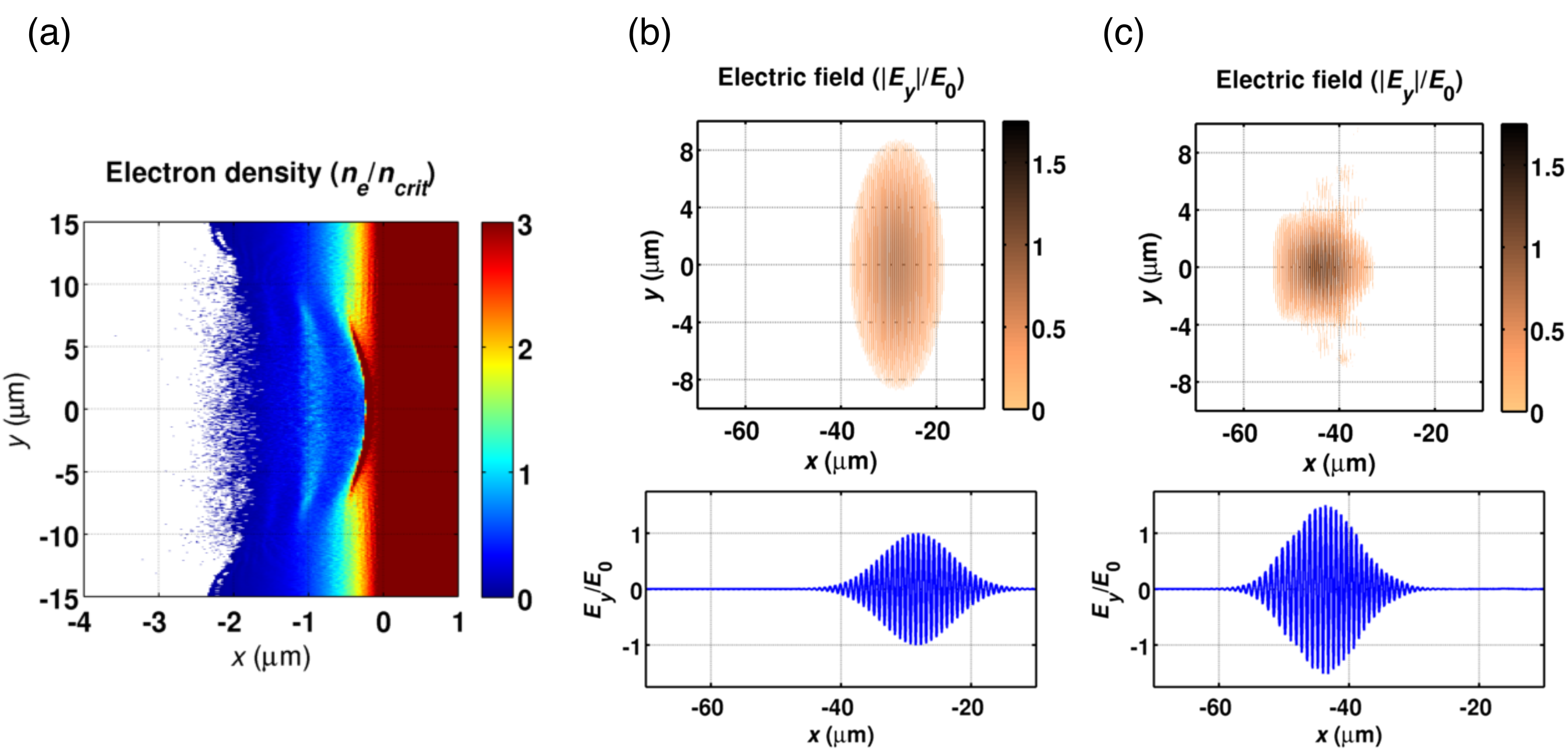}
\caption{Simulations of PM excited at relativistic intensity.  \textbf{(a)} Simulation: a snapshot of the electron density shortly after the pulse has been reflected. The pulse is less than 20 $\mu$m away from the mirror at this point.  \textbf{(b)}Simulation: a snapshot of the ($x,y$) and on-axis $y$-component of the electric field ($E_y$) in the incoming and \textbf{(c)} reflected beam at distances less than 50 $\mu$m from the plasma surface. The electron density is normalized to the critical density, $n_{crit} \equiv m \omega^2 / 4 \pi e^2$, and the electric field is normalized to $E_0 \equiv a_0 m \omega c / e$. The normalized vector potential in this run is $a_0 = 1.52$ ($I = 5 \times 10^{18}$ W/cm$^{-2}$).
 }
\label{figure_ref_sim}
\end{figure*}
 
To help understand the PM reflectivity results in Fig.~6, we performed a set of two-dimensional simulations using the multi-dimensional, fully electromagnetic, relativistic PIC code EPOCH\cite{EPOCH}.  In the simulations, a Gaussian laser pulse with its electric field in the ($x,y$)-plane, propagated along the $x$-axis. The beam size was 11 $\mu$m (FWHM) with 30 fs pulse duration (FWHM). The simulation box in the $(x,y)$-plane was 220 $\mu$m by 100 $\mu$m, with 11000 by 5000 cells, respectively.  To simplify the simulation, we assumed a pre-ionized mirror with its unperturbed surface at $x = x_0$, and electron density for $x > x_0$ set at 10 $n_{\rm{crit}}$, the ionization levels that would be achieved due to tunneling ionization in a laser field with intensity of $5 \times 10^{18}$ W/cm$^2$.  Here $n_{\rm{crit}}$ is the classical critical density for an electromagnetic wave with $\lambda = 800 nm$. A preplasma layer of thickness $l$ at the front surface of the PM was introduced to mimic the interaction of a pre-pulse with the PM.  The preplasma electron density profile for $x \leq x_0$ is $n_e =10 n_{\rm{crit}} \exp[ (x - x_0)/l]$.  No additional ionization took place during the PIC simulation.

The blue curves in Fig.~6 shows the results of the PIC simulations for preplasmas with $l = 1.0$ $\mu$m (blue solid curve) and $l = 0.5$ $\mu$m (blue dashed curve). We calculated reflectivity by taking the ratio of total reflected energy in the laser pulse to the total incoming energy (as in the space-integrated reflectivity measurement).  Absorption of laser energy by the PM is the only source of reflectivity decrease in the simulation.   For sub-relativistic intensity ($I \alt 5\times10^{17}$ W/cm$^2$), both simulations yield $\sim100\%$ reflectivity.  Thus the measured $80\%$ spatially-integrated reflectivity must be attributed to $20\%$ of the pulse energy residing in non-Gaussian wings or side lobes that fail to create an overdense plasma.  For relativistic intensity, the pulse can transfer energy into longitudinal electron motion in an amount determined by the interaction length (see Ref. ~\citen{arefiev2014}  for a corresponding discussion and references therein). Thus the reflectivity drops more for preplasma with $l = 1.0 \mu$m than for one with $l = 0.5$ $\mu$m as $I$ becomes relativistic.  The simulation with $l = 1.0$ $\mu$m yields $\sim25\%$ absorption (\textit{i.e.} reflectivity drop) at $I = 5 \times 10^{18}$ W/cm$^2$, in good agreement with the spatially-integrated reflectivity measurement.  This curve can therefore be used to represent our PM reflectivity for further modeling, as discussed below.   However, no reasonable simulation parameters reproduce the dramatic reflectivity drop observed with the spatially-resolved measurement.  We therefore conclude that its cause is something other than the absorption.
% Fig. 4(a) shows the results of the PIC simulations for two different preplasmas with $l = 1.0$ $\mu$m (blue solid line) and $l = 0.5$ $\mu$m (blue dashed line). We calculate the reflectivity by taking a ratio of the total reflected energy in the laser beam to the total incoming energy. The decrease of the reflectivity in this case is only due to the absorption of the laser energy by PM during the reflection, since we collect all of the reflected energy. The absorption in  the presence of the preplasma increases at relativistic intensities as the electron motion qualitatively changes at these intensities. Energy can be transferred into longitudinal electron motion by a pulse of relativistic intensity. The amount is determined by the longitudinal interaction length, which explains why the reflectivity drops more for the preplasma with $l = 1.0$ $\mu$m than for the preplasma with $l = 0.5$ $\mu$m. We find that even for $l = 1.0$ $\mu$m the absorption accounts for less than 30\%. This number is in agreement with the reflectivity measured using the first method. Therefore, we can conclude that the cause for the dramatic drop in the reflectivity determined using the second method is other than the absorption.

To pinpoint the cause, we examine the simulated PM surface immediately after a pulse with $I = 5 \times 10^{18}$ W/cm$^{2}$ has reflected from it.  %and a preplasma with $l = 0.5$ $\mu$m. 
Fig.~7(a) shows a snapshot of the electron density when the reflected pulse is less than 20 $\mu$m away from the mirror.  Transverse light pressure variation in the incoming beam produced a concave plasma surface with $\sim 40\mu$m radius of curvature. This curved plasma mirror focuses the reflected pulse in the near-field region, as confirmed by its widened transverse $k$-spectrum.  Thus in the far field, the reflected pulse can diverge outside of the aperture of narrow collecting optics. We therefore attribute the dramatic drop in the narrow ($f/20$) cone reflectivity results to the relativistically curved PM surface rather than absorption.

The above analysis, together with measured e-beam charge [Fig.~4(a)], enables calculation of x-ray photon number based on theoretical work of Ref.~\citen{Catravas01}. We take $a_0$ of the backscattering pulse to be $a_0$ at the exit of the LPA [red dot-dashed curve in Fig.~5(b)] multiplied by reflectivity of the PM, for which we use the blue solid curve in Fig.~6. The result, plotted as a function of $\bar{n}_e$, is the black solid curve in Fig.~5(a) and ranges from 1.2 to 0.7 within the quasi-monoenergetic tuning range of the e-beam.  This corresponds to intensity 4 to 10 times higher than achieved with CBS by a separate counter-propagating laser pulse \cite{Power14}.  The blue dashed line in Fig.~4(b) then shows the calculated x-ray photon number vs.~central x-ray photon energy.  The calculated and measured photon numbers, averaged over x-ray photon energy, are $2.7\times 10^{7}$ and $2.0\times 10^{7}$, respectively.  The calculated values are within the the error bars of the measured values, and confirm the weak dependence of photon number on photon energy. This good agreement confirms the analysis of exit driver intensity and PM reflectivity, and shows that the automatic overlap between e-beam and the peak of the reflected scattering pulse is very high.  

X-ray brightness can be estimated from the measured beam divergence and photon number distribution, assuming a 6 $\mu$m source size \cite{Chen13} and 30 fs x-ray pulse.  The result is $10^{19}$photons s$^{-1}$ mm$^{-2}$ mrad$^{-2}$ (per $0.1\%$ bandwidth) for 190 KeV x-rays.  Energy conversion efficiency from laser pulse to X-ray beam is $\sim 10^{-6}$, while photon conversion efficiency is $\sim 6\times 10^{-12}$, the highest yet achieved for LPA-based mono-energetic Compton sources.

Our analysis suggests two opportunities for further improving x-ray conversion efficiency and brightness in future work.  First, sharpening the density down-ramp at the gas jet exit can potentially improve the transmitted $a_0$ by as much as a factor of two [see black dashed and red curves in Fig.~5(b)], thus quadrupling the intensity of the backscattering pulse\cite{Palastro2014}.  Second, sharper relativistic curvature of the PM surface [see Fig.~7(a)] by a more intense transmitted drive pulse could potentially focus the retro-reflected light onto oncoming electrons.  As an illustration of this effect for conditions of the present experiments, the light green contours in Figs.~7(b)-(c) show snapshots of the ($x,y$) and $y$-component of the electric field ($E_y$) in the incoming and reflected beam, respectively, $\sim 45 \mu$m in front the PM surface, where the tightest focus occurs. The reflected beam is visibly more focused due to the curved PM surface, while $E_y$ is $\sim 1.7\times$ higher [see blue curve in Fig.~7(c), compared to 7(b)], corresponding to 3-fold intensity increase.  Unfortunately, for current conditions, the electron bunch is only $\sim10 \mu$m behind the laser driver, so backscatter occurs only $\sim 5 \mu$m from the PM surface, where the intensity enhancement is negligible.  Nevertheless, this enhancement could become significant in LPA experiments at lower $\bar{n}_e$, in which the e-bunch propagates further behind the driver due to larger bubble size, or when the transmitted drive pulse has higher $a_0$.  
 
%The observed field enhancement can be beneficial in the context of the CBS. However, the timing of the enhancement and its location have to be such that the electron bunch that follows the laser driver can sample it. If the electron bunch lags behind the driver by a distance $l_e$, then the focusing has to occur not further than $l_e$ away from the mirror in order to be beneficial for CBS. In our experiments, Therefore, the enhancement that occurs roughly 45 $\mu$m from the plasma mirror in the PIC simulations should not impact our CBS results. Optimization of the laser parameters are needed to fully benefit from the induced curvature of the plasma mirror.

%%%%%%%%%%%%%%%%%%%%%%%%%%%%%%%%%%%%%%%%%%%%%%%%%%%%%%%%%%%%%%
%%%%%%%%%%%%%%%%   SECTION V           %%%%%%%%%%%%%%%%%%%%%%%%%%%%%%%%%%%

\section{CONCLUSION}

In conclusion, we demonstrated quasi-monoenergetic Compton backscatter X-ray generation using the easily aligned combination of an LPA with a single drive pulse and a PM. The central x-ray photon energy was tuned from 75 KeV to $\sim$200 KeV, and can be scaled to MeV energy
by tuning up e-beam central energy. The Compton source has photon number $2 \times 10^{7}$, divergence $\sim10$ mrad, and brightness $10^{19}$photons s$^{-1}$ mm$^{-2}$ mrad$^{-2}$ (per $0.1\%$ bandwidth).  Drive pulse transmission through the LPA and PM reflectivity were fully characterized by measurement and simulation, yielding a complete quantitative understanding of CBS x-ray properties. 

Our analysis suggest that x-ray brightness could be increased as much as 10-fold In future experiments by sharpening the density down-ramp at the gas jet exit, and optimizing PM focus at relativistic intensity. Such improvements might also enable study of nonlinear Compton backscatter, which requires a strongly relativistic scattering pulse.

%%%%%%%%%%%%%%%%%%%%%%%%%%%%%%%%%%%%%%%%%%%%%%%%%%%%%%%%%%%%%%
%%%%%%%%%%%%%%%%   SECTION VI           %%%%%%%%%%%%%%%%%%%%%%%%%%%%%%%%%%%

\section{ACKNOWLEDGEMENT}
Experimental work was supported by DOE grants DE-SC0012444 and DE-SC0011617, AFOSR grant FA9550-14-1-0045, and Robert Welch Foundation grant F-1038. H.-E. T. acknowledges support from NSF grant PHY-1354531.  3D PIC modeling was supported by DOE contracts DE-SC0007889 and DE-SC0010622. Plasma mirror simulations were performed by A. V. A. using the EPOCH code (developed under UK EPSRC grants EP/G054940/1, EP/G055165/1 and EP/G056803/1) and HPC resources provided by the Texas Advanced Computing Center. A. V. A.  was supported by AFOSR Contract FA9550-14-1-0045, NNSA Contract DE-FC52-08NA28512, and DOE Contract DE-FG02-04ER54742.

%%%%%%%%%%%%%%%%%%%%%%%%%%%%%%%%%%%%%%%%%%%%%%%%%%%%%%%%%%%%%%
%%%%%%%%%%%%%%%%   REFERENCE          %%%%%%%%%%%%%%%%%%%%%%%%%%%%%%%%%%%

\end{document}